\documentstyle[prl,aps,twocolumn,epsf]{revtex}

\begin{document}

\title{\bf A Nonequilibrium Field Theory Description of the 
Bose-Einstein Condensate}

\author{D. G. Barci$^{a,c}$, E. S. Fraga$^b$ and Rudnei O. Ramos$^c$}
\address{  (a) {\it Department of Physics, University of Illinois at 
Urbana-Champaign, 1110
W. Green St., Urbana, IL  61801-3080, USA} \newline
(b){\it Physics Department, 
Brookhaven National Laboratory, Upton, NY 11973, USA} \newline
(c) {\it Universidade do Estado do Rio de Janeiro,
Instituto de F\'{\i}sica, Departamento de F\'{\i}sica Te\'orica,
 20550-013 Rio de Janeiro, RJ, Brazil}}

\maketitle

\begin{abstract}
We study the detailed out of equilibrium time evolution of a homogeneous 
Bose-Einstein condensate.
We consider a nonrelativistic quantum theory for a self-interacting complex 
scalar field, immersed in a thermal bath,
as an effective microscopic model for the description of the 
Bose-Einstein condensate. This approach yields the 
following main results: (i) the interaction between fluctuations proves 
to be crucial in the mechanism of instability generation; (ii) there are 
essentially two regimes in the $k$-space, 
with a crossover for $k^2/2m \sim 2\lambda |\varphi_0|^2$, 
where, in our notation, $\lambda$
is the coupling constant and $|\varphi_0|^2$ is the condensate density; 
(iii) a set of coupled equations that determines 
completely the nonequilibrium dynamics of the condensate density as a 
function of the temperature and of the total density of the gas.
%
%
\noindent
PACS number(s): 03.75.Fi, 05.30.Jp, 11.10.Wx
\end{abstract}


\bigskip

The experimental verification of the phenomenon of Bose-Einstein
condensation in weakly interacting gases 
has boosted a large number of theoretical investigations on the dynamics
of weakly-interacting dilute gas systems [for a recent review, see e.g.
Ref. \cite{trento} and references therein]. Current experiments
and planned ones make it possible to probe different aspects
of the Bose-Einstein condensate formation, with great control over 
interactions, trapping potentials, etc. Nevertheless, a basic problem not yet 
fully understood is the following: given an initial state, how will 
the condensate evolve with time? In special, the 
time scales for the condensate formation and its final size are important 
quantities involved in recent experiments with dilute atomic 
gases \cite{growth}.

On the theoretical side, however, only restrict progress has been achieved 
concerning the problems above. Previous studies by Stoof 
\cite{stoof} were able to  give a qualitative idea of the various
time scales involved during the condensate formation. In fact, they were the 
first attempts to analyze the problem from a microscopic point of view, by
using the Schwinger-Keldysh closed time-path formalism 
(for reviews, see for instance Refs. \cite{keldysh,lebellac} )
in the quantum field theory description of Bose-Einstein condensation.
Regarding the condensate growth problem, Gardiner {\it et al.}
\cite{gardiner} have used a quantum kinetic theory to construct a master 
equation for a density operator describing the state of the condensate, which
is equivalent to a Boltzmann equation describing a quasi-equilibrium 
growth of the condensate.     

In this work we will study the quantum field 
time evolution of an interacting homogeneous condensate. Although 
non-homogeneity is inherent to current experiments on Bose-Einstein 
condensation of atomic gases in trapping potentials, 
we believe that a full understanding of the time evolution of
even the simpler case of a homogeneous gas is still lacking. 
Besides, as pointed out by Stoof in \cite{stoof}, the simplest
formulations based on kinetic theory do not allow for the
observation of a macroscopic occupation of the one-particle ground
state, and the question of the instability of the Bose gas system in the 
homogeneous case is a nontrivial one. This makes its study an interesting 
problem, which may shed some light on the analysis of the systems 
under experimental investigation. 

We consider the simplest model for a 
nonrelativistic complex Bose field, with a hard core interaction 
potential, whose Lagrangian density is given by 
(throughout this work we use units such that $\hbar =1$)
\begin{equation}
{\cal L}=\phi^*\left( i\frac{d}{dt}+\frac{1}{2m}\nabla^2\right)\phi
+\mu \phi^* \phi -\lambda(\phi^*\phi)^2 \; ,
\label{eq1}
\end{equation}
where the complex scalar field $\phi({\bf x},t)$ represents complex spinless 
bosons of mass $m$, and $\lambda$ is the coupling constant, related 
to the $s$-wave scattering length $a$ by: $\lambda=4 \pi a/m$. 
In (\ref{eq1}) 
we have also explicitly introduced a 
chemical potential $\mu$ that produces a  constant total density of 
particles $\langle\phi^*\phi\rangle =n$.
We also assume that the system is coupled to a heat bath environment 
with inverse temperature $\beta=1/k_B T$.

We may now perform the standard decomposition \cite{beliaev} 
of the fields ($\phi,\phi^*$) into
a condensate (uniform) part ($\varphi_0,\varphi_0^*$)
and a fluctuation (nonuniform)
part ($\varphi,\varphi^*$) that describes the atoms outside the
condensate, as $\phi({\bf x},t)= \varphi_0 (t) +
\varphi ({\bf x},t)$ and  $\phi^* ({\bf x},t)= \varphi_0^* (t) +
\varphi^* ({\bf x},t)$, where we have assumed a homogeneous condensate.
Note that we take at first $\varphi_0(t)$ as an arbitrary function of time 
that will be determined by the dynamics of the system.

Substituting the fields above in (\ref{eq1}), we can readily obtain
the Bogoliubov spectrum for quasi-particles \cite{bogoliubov}. In particular, 
the quadratic part of the Lagrangian for the fluctuation fields is
the Bogoliubov approximation for quasiparticles. However, this 
approximation has a flaw: there is no interaction between
the fluctuations, which is only reasonable at temperatures
well below the critical temperature for the condensate formation.
Our aim, in this work, is the study of the condensate evolution
and, therefore, we must go beyond the Bogoliubov approximation.
The simplest extension is to implement a mean field approximation in the 
interactions between the fluctuation field. 
In this way, as we show below, one can make clear the appearance of 
instability modes towards the condensation formation once
the interactions between fluctuations are taken into account.  
With the decomposition above, the interaction term for
fluctuations in the Lagrangian becomes $\lambda (\varphi^* \varphi)^2$.
The mean-field approximation 
amounts to the following:
\begin{equation}
\!\lambda (\varphi^* \varphi)^2=\!
4 \lambda \langle \varphi^* \varphi \rangle
\varphi^* \varphi + \left[ \lambda(\varphi^* \varphi)^2 -
4 \lambda \langle \varphi^* \varphi \rangle
\varphi^* \varphi \right],
\label{meanfield}
\end{equation}
where the first term in the rhs is taken as part of the 
quadratic Lagrangian for fluctuations, and the term inside
the square brackets is taken as part of the interaction 
Lagrangian.

{}From the decomposition of the fields and
(\ref{meanfield}), the quadratic part of the Lagrangian density for
the fluctuations, ${\cal L}_0 (\varphi,\varphi^*)$, may be written as
\begin{eqnarray}
{\cal L}_0 (\varphi,\varphi^*) &=& \varphi^*\left[ i\frac{d}{dt}+\frac{1}{2m}
\nabla^2 \right]\varphi
+\varphi^* (-\lambda \varphi_0^2 ) \varphi^* \nonumber \\
&+& \varphi (-\lambda {\varphi_0^*}^2 ) \varphi \; .
\label{eq2}
\end{eqnarray}
Here, we have used the fact that, under the field decomposition in 
the condensate and out of the condensate modes, the density constraint 
then becomes $\langle \phi^* \phi \rangle =
|\varphi_0|^2 + \langle \varphi^* \varphi \rangle = n$.
Additionally, assuming that at the initial time the system is mostly composed
of particles outside the condensate, $\langle \phi^* \phi \rangle \simeq
\langle \varphi^* \varphi \rangle$ (at $t=0$), simple relations
involving the generating functional for the correlation functions
(see, for instance, the last section of chapter 2 in \cite{baym})
allow us to write the total number density 
$n$ of particles in terms of the chemical potential $\mu$, valid in the
mean-field approximation for the potential, as:
$\mu = 4 \lambda n$. Note that this is just the expression obtained
also in the Hartree-Popov approximation \cite{griffin2}, which
turns out to satisfy the Hugenholtz-Pines
relation \cite{pines} that would be obtained in the 
equilibrium problem. These considerations lead to the quadratic
Lagrangian for the fluctuations shown above. 

The scenario we have in mind is that for time $t<0$ the initial state 
is in {\it equilibrium} at a temperature $T_i \gg T_c$. At
$t=0$ the system is then abruptly quenched
to a much lower temperature
$T_f \ll T_c$. $T_f$ is the temperature of the thermal bath
in which the system is immersed and, of course, it will be the
equilibrium temperature which the system will reach asymptotically. 
This kind of quench is easily attained in the experiments
of Bose-Einstein condensation of atomic gases, where the typical
relaxation time scales are long enough (around $\sim 0.1 s$,
depending on the temperature \cite{MIT}) to allow for a fast drop
in the temperature of the system that evolves afterwards
out of equilibrium. 
With
this choice of initial state, it is reasonable to approximate the
dynamics of the build up of the condensate state, which at
the initial time is $n_{\rm cond.}(t=0) = |\varphi_0(t=0)|^2 \approx
0$,
and the depletion of the excited states (which at $t=0$ it is
given by $n \approx n_{\rm exc.}(t=0) =\langle
\varphi^* \varphi \rangle$) as essentially a two-level 
problem. It is clear that this approximation breaks down
for temperatures close to the critical temperature, where
the detailed treatment would require a thorough study
of the dynamics among the many levels of excited states.
In the above approximation, the condensate builds up
subject to the density constraint relation, which may be
expressed in terms of the averages of the real and the imaginary
parts of $\varphi$ and $\varphi^*$
($\varphi = \varphi_1 + i \varphi_2$ and $\varphi^* 
= \varphi_1 - i \varphi_2$, respectively). Spatial translational 
invariance yields:
\begin{eqnarray}
\lefteqn{|\varphi_0 (t)|^2 + n_{\rm exc.}(t) = n \;,} \nonumber \\
& & n_{\rm exc.}(t) = \langle
\varphi_1 (t) \varphi_1 (t) \rangle + \langle
\varphi_2 (t) \varphi_2 (t) \rangle \;.
\label{dens2}
\end{eqnarray}
The field averages above can be expressed in terms of the
Green's functions for $\varphi_1$ and $\varphi_2$ as ($j=1,2$)
\begin{equation}
\langle \varphi_j (t) \varphi_j (t) \rangle = \int \frac{d^3 k}{(2 \pi)^3}
\left[-i G_{jj}^> ({\bf k},t,t) \right] \;,
\label{average}
\end{equation}
where $G_{jj}^> ({\bf k},t,t)$ is defined from the Green's functions
for the fields in the closed-time path \cite{stoof,lebellac} 
(in momentum space)
\begin{eqnarray}
&&G_{jj}^{++} ({\bf k},t,t')  =  G_{jj}^> ({\bf k},t,t') \theta(t-t')+
G_{jj}^< ({\bf k},t,t') \theta(t'-t)\; , \nonumber \\
& & G_{jj}^{--} ({\bf k},t,t')  =  G_{jj}^> ({\bf k},t,t') \theta(t'-t)+
G_{jj}^< ({\bf k},t,t') \theta(t-t')\; , \nonumber \\
& & G_{jj}^{+-} ({\bf k},t,t')  =  -G_{jj}^< ({\bf k},t,t')\; , \nonumber \\
& & G_{jj}^{-+} ({\bf k},t,t')  =  -G_{jj}^> ({\bf k},t,t')\;.
\label{greens}
\end{eqnarray}
The functions  $G^>$ and $G^<$ satisfy the property
$G^<_{jj} ({\bf k},t,t') = 
G^>_{jj} ({\bf k},t-i \beta,t')$, which is recognized as the
periodicity condition in imaginary time (Kubo-Martin-Schwinger (KMS)
condition). $\beta$ here is the inverse of the temperature of the
thermal bath and appears here as a consequence of the  
boundary conditions arising from the construction of the complex time 
path.
$G^>$ and $G^<$ are constructed from the homogeneous solutions
to the operator of quadratic fluctuation modes, which, using Eq. (\ref{eq2})
expressed in terms of $\varphi_1$ and $\varphi_2$, are given by
(in momentum space)
\begin{eqnarray}
\lefteqn{\frac{d \chi_2 ({\bf k},t)}{d t} + \left(\frac{{\bf k}^2}{2 m} + 
2 \lambda |\varphi_0|^2 \right) \chi_1 ({\bf k},t)= 0}  \nonumber \\
& & \frac{d \chi_1({\bf k},t)}{d t} - \left(\frac{{\bf k}^2}{2 m} - 
2 \lambda |\varphi_0|^2 \right) \chi_2 ({\bf k},t)= 0 \;,
\label{modes}
\end{eqnarray}

The boundary conditions for the solutions of the equations 
above are such that, for $t <0 $, $|\varphi_0 (t)|^2=0$,
$\chi_1 ({\bf k},t) = \cos(\varepsilon_{\bf k}t)$ and 
$ \chi_2 ({\bf k},t)=- \sin(\varepsilon_{\bf k}t)$, where 
$\varepsilon_{\bf k}={\bf k}^2/(2 m)$.
In terms of these fluctuations modes, the Green's functions
are expressed as

\begin{eqnarray}
\lefteqn{G^>_{jj}({\bf k},t,t') = \frac{i}{2 \left(1-e^{-\beta 
\varepsilon_{\bf k}} \right)}} \nonumber \\
& &\times \left[ 
\chi_j ({\bf k},t)\chi_j^* ({\bf k},t')
+ e^{-\beta \varepsilon_{\bf k}}\chi_j^* ({\bf k},t)\chi_j ({\bf k},t') 
\right]
\label{G>}
\end{eqnarray}

\noindent
and $G^<_{jj} ({\bf k},t,t')=G^>_{jj}  ({\bf k},t',t)$.

By decoupling the set of equations in (\ref{modes}), we can readily
identify that those modes with $({\bf k}^2/2m) < 2 \lambda 
|\varphi_0|^2$ are unstable and drive the excited particles
towards condensation. Note also that not all the excited
particles condense, since there will always be a fraction
(which depends on various parameters for a particular
system and on the temperature of the thermal bath)
of excited modes, with high enough frequency, that 
remains stable. This will be clear from our numerical results 
shown later.
 
Using Eqs. (\ref{modes}), (\ref{G>}) and the boundary conditions on the
Green's functions, Eq. (\ref{greens}),
together with the initial condition on the density (at $t=0$,
as defined before), one can then
show that $ n_{\rm exc.}(t)$ can be expressed as (by subtracting the
zero point divergent contribution)
\begin{equation}
n_{\rm exc.}(t) \!=\!
\left(\frac{\beta}{\beta_c}\right)^{3/2}\!\!\!
\int \! \frac{d^3 k}{(2 \pi)^3} \! \! \left[|\chi_1({\bf k},t)|^2
+ |\chi_2 ({\bf k},t)|^2 \right] n_{{\bf k}}(\beta) ,
\label{integral}
\end{equation}
where $n_{{\bf k}}(\beta)=(e^{\beta\epsilon_{{\bf k}}}-1)^{-1}$,
and $\beta_c$ is the inverse of the equilibrium critical temperature,
defined in terms of the total gas density, $n$ \cite{griffin}
It should also be noted that in our out of 
equilibrium approach there are no infrared divergences since the finite time 
is 
a natural regulator. However for the equilibrium $t\rightarrow\infty$,
the critical temperature 
will be modified by the interactions as pointed out in Ref.
\cite{GB}.
   
The expression above can also be obtained
directly in terms of the Green's functions for the complex
fields $\varphi,\varphi^*$:
\begin{equation}
\langle \varphi (t) \varphi^* (t) \rangle = \int \frac{d^3 k}{(2 \pi)^3}
\left[-i G_{\varphi \varphi^*}^> ({\bf k},t,t) \right]\;,
\label{averagec1}
\end{equation}
\begin{equation}
\langle \varphi^* (t) \varphi (t) \rangle = \int \frac{d^3 k}{(2 \pi)^3}
\left[-i G_{\varphi^* \varphi}^> ({\bf k},t,t) \right] \;.
\label{averagec2}
\end{equation}
In terms of (\ref{averagec1}) and (\ref{averagec2}), we have 
$ \langle
\varphi_1 (t) \varphi_1 (t) \rangle + \langle
\varphi_2 (t) \varphi_2 (t) \rangle = (\langle
\varphi (t) \varphi^* (t) \rangle + \langle
\varphi^* (t) \varphi (t) \rangle)/2$, and the KMS condition can be 
expressed, in this case, as $G_{\varphi \varphi^*}^> ({\bf k},t-i \beta,t')
=[G_{\varphi^* \varphi}^> ({\bf k},t,t')]^*$, or 
$G_{\varphi^* \varphi}^> ({\bf k},t-i \beta,t')
=[G_{\varphi \varphi^*}^> ({\bf k},t,t')]^*$. 

Equation (\ref{integral}) is the first order term in the finite temperature
quantum many-body perturbation expansion for $\langle \varphi_j^2 \rangle$.
Higher-order corrections for the equal-time two-point field
averages can be  expressed
in terms of the coincidence limit of the (causal)
two-point Green's functions $G_{\varphi \varphi^*}$ and 
$G_{\varphi^* \varphi}$, which satisfy
the Dyson equations (the indices stand for $\varphi$ and $\varphi^*$):
\begin{equation}
G_{ij}=G_{ij}^0 + G_{ik}^0\Sigma_{kl} G_{lj}\;,
\label{fullG}
\end{equation}
where $\Sigma_{ij}$ is the (matrix) self-energy, and $G_{ij}^0$ is the
zeroth-order non-interacting Green's function,
satisfying the equation (in momentum space) 
\begin{equation}
\left[ \pm i \frac{d}{dt} - \varepsilon_{\bf k} \right] 
G^0_{\varphi \varphi^*(\varphi^* \varphi)}({\bf k},t,t') = \delta(t-t)\;.
\label{G0}
\end{equation}
One of the advantages of expressing the Green's functions in terms
of the solutions of (\ref{modes}) is the possibility of obtaining, 
in an unambiguous way, all higher-order corrections to the two-point
and many-point functions \cite{nextwork}.

By using Eq. (\ref{integral}), we can rewrite the constraint on the density as 
\begin{eqnarray}
|\varphi_0(t)|^2&=&\frac{(\beta/\beta_c)^{3/2}}{2\pi^2}
\int_0^{\sqrt{16\pi a|\varphi_0(t)|^2}}
dk~k^2~\times\nonumber \\
&\times& \left[ 1-
\left(|\chi_1({\bf k},t)|^2 + |\chi_2({\bf k},t)|^2\right)\right]
n_{{\bf k}}(\beta) \; ,
\label{vinculo}
\end{eqnarray}
that reproduces the result obtained by Stoof \cite{stoof} for the limit 
$t\to\infty$. Equations (\ref{modes}) and (\ref{vinculo}) form an 
integro-differential system that may be solved 
for $\varphi_0(t)$ numerically, given the initial conditions for 
$\varphi_0(t)$, 
$\chi_1(\vec{k},t)$ and $\chi_2(\vec{k},t)$ mentioned before. Indeed, this 
system of equations determines completely the time evolution of the 
condensate density as a function of the temperature and of the total density 
of the gas. Explicit results for different temperatures are shown in {}Fig. 1.
It is important to point out at this stage that the evolution of the 
condensate is 
completely driven by the interactions between the microscopic fluctuations
of the field around the condensate.  

Throughout this letter, we have developed an out-of-equilibrium 
non-perturbative quantum field theory description of the condensation 
process of an interacting homogeneous Bose-Einstein gas quenched below 
the critical temperature. In summary, this approach yielded the 
following main results: (i) The interaction between fluctuations proved 
to be crucial in the mechanism of instability generation; without it, 
there is simply no macroscopic condensate at all. (ii) There are essentially  
two regimes in the $k$-space: for $(k^2/2m)<<2\lambda|\varphi_0^2|$, we have
unstable modes that decay exponentially, while for
$(k^2/2m)>>2\lambda|\varphi_0^2|$, we have stable modes that oscillate,
with a crossover for $(k^2/2m) \sim 2\lambda|\varphi_0^2|$. (iii) Equations (\ref{modes}) and 
(\ref{vinculo}) come from a microscopic model for the weak-interaction gas, 
and determine completely the dynamics of the condensate. In fact, they are 
non-perturbative and 
certainly implement a resummation of the ladder Feynman diagrams mentioned by 
Stoof in \cite{stoof}. Indeed, the highly nonequilibrium character  of 
this description should complement the usual approach via Boltzmann 
equation.

Although we focused this letter on the instability process that generates 
the condensate ({\it i.e.}, the short time behavior), for $t\rightarrow\infty$ 
our 
results confirm the behavior predicted in Ref. \cite{stoof} for this limit. 
However, the equilibrium 
($t\rightarrow\infty$) values of the condensate fraction are lower 
than the experimental results \cite{MIT} and the calculations of 
Dalfovo {\it et. al.} \cite{trento}. 
This may be due to our approximation of neglecting
incoherent
collisional processes, which is a valid approximation in an infinite 
homogeneous gas at very low temperatures and densities 
but otherwise may give an important contribution. We expect that the
self-consistent inclusion of pair terms should account for most
of these contributions. We will report on these improvements of our
mean-field approximation in a future publication \cite{nextwork}.

In spite of the absence of non-homogeneity effects, we hope 
that the approach developed here may be useful in the analysis of 
transients in realistic Bose-Einstein condensation experiments with atomic 
gases. 
Moreover, with a suitable generalization of the formalism presented above, 
we could be able to develop a theoretical description of the 
dynamical aspects of a recently proposed experiment \cite{chiao} 
(currently in progress), regarding the Bose-Einstein condensation in a 
weakly-interacting photon gas in a nonlinear {}Fabry-Perot cavity 
\cite{alessandro}.

\begin{figure}[c]
\epsfysize=9 cm 
{\centerline{\epsfbox{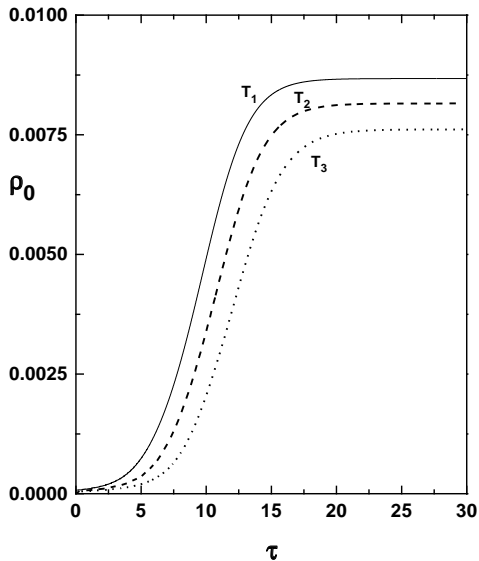}}}
\caption{Condensate density as a function of time for $na^3=0.01$ and
$T_1/T_c=0.06$, $T_2/T_c=0.08$ and $T_3/T_c=0.1$. Here, 
$\tau\equiv (\hbar/ma^2)t$ is a dimensionless time and 
$\rho_0 \equiv a^3 |\varphi_0|^2$ is a dimensionless density. }
\end{figure}


\acknowledgements

It is a pleasure to thank D. Boyanovsky, E. Fradkin, S.P. Sorella, 
A. Tanzini and
S.E. Jor\'as for fruitful conversations. We also thank 
R.Y. Chiao for interesting discussions regarding the dynamics 
of the photon fluid. The authors acknowledge CNPq and FAPERJ  
for financial support. 
E.S.F thanks the hospitality of DFT at UERJ, where this work has
started.
D.G.B. is also partially supported by NSF 
through grant DMR-9817941. 
E.S.F. is also partially supported by the U. S. Department of Energy under 
Contract No. DE-AC02-98CH10886.



\begin{references}


\bibitem{trento} F. Dalfovo, S. Giorgini, L. P. Pitaevskii and S. Stringari,
Rev. Mod. Phys. {\bf 71}, 463 (1999).

\bibitem{growth} H.-J. Miesner, {\it et al.}, Science {\bf 279}, 1005 (1998).

\bibitem{stoof} H. T. C. Stoof, J. Low Temp. Phys. {\bf 114}, 11 (1999).
Phys. Rev. Lett. {\bf 66}, 3148; Phys. Rev. {\bf A49}, 3824 (1994); 
Phys. Rev. {\bf A45}, 8398 (1992).

\bibitem{keldysh} K. Chou, Z. Su, B. Hao and L. Yu, Phys. Rep. {\bf 118}, 1
(1985).

\bibitem{lebellac} M. Le Bellac {\it Thermal Field Theory} 
(Cambridge University Press, Cambridge, 1996).

\bibitem{gardiner} C. W. Gardiner, P. Zoller, R. J. Ballagh and M. J. Davis,
Phys. Rev. Lett. {\bf 79}, 1793 (1997); C. W. Gardiner, M. D. Lee, 
R. J. Ballagh, M. J. Davis and P. Zoller, Phys. Rev. Lett. {\bf 81}, 
5266 (1998).

\bibitem{beliaev} S. T. Beliaev, Sov. Phys. JETP {\bf 7}, 289 (1958);

\bibitem{bogoliubov} A. A. Abrikosov, L. P. Gorkov, and I. E. 
Dzyaloshinski, {\it Methods of Quantum Field Theory in Statistical 
Physics} (Dover Publ., Inc., New York, 1975).

\bibitem{baym} L. P. Kadanoff and G. Baym, {\it Quantum Statistical
Mechanics, Green's Functions Methods in Equilibrium and Nonequilibrium
Problems} (Addison-Wesley Publ. Co., New York, 1989).

\bibitem{griffin2} A. Griffin, Phys. Rev. {\bf B53}, 9341 (1996).

\bibitem{pines} N. Hugenholtz and D. Pines, Phys. Rev. {\bf 116},
489 (1959).

\bibitem{MIT} M.-O. Mewes, M. R. Anderson, N. J. van Druten, 
D. M. Kurn, D. S. Durfee, C. G. Townsend and W. Ketterle, 
Phys. Rev Lett. {\bf77}, 988 (1996).

\bibitem{griffin} A. Griffin, {\it Excitations in a Bose-condensed liquid}
(Cambridge University Press, Cambridge, 1993).

\bibitem{GB} G. Baym, {et al.}, Phys. Rev. Lett. {\bf 83}, 1703 (1999). 

\bibitem{nextwork} D. G. Barci, E. S. Fraga and R. O. Ramos, work
in progress.

\bibitem{chiao} R. Y. Chiao and J. Boyce, Phys. Rev. {\bf A60}, 
4114 (1999).

\bibitem{alessandro} A. Tanzini and S. P. Sorella, 
Phys. Lett. {\bf A263}, 43 (1999).


\end{references}
\end{document}